\documentclass[amsmath,amssymb]{revtex4}
\usepackage{txfonts}%
\usepackage{graphicx}
\usepackage{epsfig}
\usepackage{graphics,color}
\usepackage{amssymb}
\usepackage{amsbsy}

 \newcommand{\be}{\begin{equation}}
 \newcommand{\ee}{\end{equation}}
 \newcommand{\bse}{\begin{subequations}}
 \newcommand{\ese}{\end{subequations}}
 \newcommand{\bea}{\begin{eqnarray}}
 \newcommand{\eea}{\end{eqnarray}}

\newcommand{\bean}{\begin{eqnarray*}}
\newcommand{\eean}{\end{eqnarray*}}

\begin{document}

\title{Dependence of elliptic flow on transverse
 momentum  in $\sqrt{\mathrm{\it s_{NN}}}=200$
 GeV Au-Au and $\sqrt{\mathrm{\it s_{NN}}}=2.76$ TeV Pb-Pb collisions} \vskip1.0cm
\author{Bao-Chun Li\footnote{libc2010@163.com, s6109@sxu.edu.cn}, Yuan-Yuan
Fu, Er-Qin Wang,  Fu-Hu Liu\footnote{fuhuliu@163.com,
liufh@mail.sxu.cn}} \affiliation{Department of Physics and Institute
of Theoretical Physics, Shanxi University, Taiyuan, Shanxi 030006,
China}

\vskip2.0cm

\begin{abstract}
We investigate the dependence of elliptic flows $v_2$ on transverse
momentum $P_T$ for charged hadrons produced in nucleus-nucleus
collisions at high energy by using a multi-source ideal gas model
which includes the interaction contribution of the emission sources.
Our calculated results are approximately in agreement with the
experimental data over a  wider $P_T$ range from the STAR¡¡and ALICE
Collaborations. It is found that the expansion factor increases
linearly with the impact parameter from most central (0-5\%) to
mid-peripheral (35-40\%) collisions.

\end{abstract}

\maketitle

{\section{Introduction}}

The first elliptic flow data from the Large Hadron Collider(LHC)
obtained by the ALICE Collaboration~\cite{Aamodt:2010pa} are similar
as observed at lower energies at the Relativistic Heavy Ion Collider
(RHIC)~ \cite{Adare:2010ux, Hirano:2010je}. The elliptic flow of
charged hadrons  as a function of transverse momentum $v_2(P_T)$
increases with $P_T$ and saturates at higher $P_T$. The similarity
between the elliptic flows at LHC and RHIC is consistent with the
predictions of the viscous hydrodynamic model~\cite{Luzum:2010ag}.
The elliptic flow is generated in a collective expansion of the
dense matter created in the collisions. The comparison to model
calculations can provide valuable information of the collisional
evolution. Recent results of relativistic viscous hydrodynamic
calculations for Pb-Pb collisions at $\sqrt{\mathrm{\it
s_{NN}}}=2.76$ shows systematic deviations from the ALICE data in
the small momenta region $P_T< 800$ MeV~\cite{Bozek:2011wa}. The
deviations is interpreted to be a non-negligible contribution of
non-thermalized particles from jet fragmentation. The origin of this
deviation is still under debate. Elliptic flow results from the
interactions among the produced particles and can be used to probe
local thermodynamic equilibrium. By fitting the experimental data of
elliptic flows, it has been found in Ref.~\cite{Song:2011qa} that
the average Quark Gluon Plasma (QGP) specific shear viscosity
$\eta/s$ slightly increases from RHIC to LHC energies in the newly
developed hybrid model VISHNU, which connects viscous hydrodynamics
with a hadron cascade model. At the  LHC energies, it remains an
open question whether the QGP fluid is more viscous or more perfect.

  Elliptic flow is a measure of the azimuthal anisotropy of
particle momentum distributions in the plane perpendicular to the
beam direction. It results from the initial spatial anisotropy in
non-central collisions and is thus sensitive to the properties of
the hot dense matter formed during the initial stage of heavy ion
collisions. In a thermodynamical picture, the asymmetric
distribution of the initial energy density causes a larger pressure
gradient in the shortest direction of the ellipsoidal medium, which
can lead to an anisotropic azimuthal emission of particles. As a
signature of collective flow in relativistic nuclear collisions, the
elliptic flow was experimentally observed at the AGS, SPS, RHIC, and
LHC,respectively~\cite{Ollitrault:1992bk,Ackermann:2000tr,Alt:2003ab,Barrette:1996rs,Barrette:1994xr}.
 In recent years, it has been believed that
the system created at the RHIC~\cite{Ackermann:2000tr} is a strongly
coupled quark-gluon plasma (sQGP) and behaves like a nearly
``perfect" fluid by using the hydrodynamic simulation to the
elliptic flow at the RHIC~\cite{Hydro,Hydro2}.\\

{\section{Multi-source ideal gas model}}

The multi-source ideal gas model is developed by
us~\cite{liu,liub,liuc} from the Maxwell's ideal gas model. In this
section, we briefly outline the calculation in the model.  Many
primary nucleon-nucleon collisions happen at the initial stage of
nucleus-nucleus collisions. Each primary nucleon-nucleon collision
can be regarded as an emission source (a compound hadron fireball)
at intermediate energy or a few sources (wounded partons and
woundless partons) at high energy. The participant nucleons that
take part in primary collisions have probabilities to take part in
cascade collisions with latter nucleons. Meanwhile, the particles
produced in primary or cascade nucleon-nucleon collisions have
probabilities to take part in secondary collisions with latter
nucleons and other particles. Each cascade (or secondary) collision
is also regarded as an emission source or a few sources. Many
emission sources of final-state particles are expected to be formed
in the collision process, and each source is considered to be a
Maxwell¡¯s ideal gas in the rest source frame. There are
interactions among the emission sources due to the mechanics and
electromagnetism effects. The interactions will cause the source to
depart from the isotropic emission. To describe the anisotropic
source, the sources deformation and movement in the model have been
introduced.

Let the beam direction be the $oz$ axis and the reaction plane be
the $xoz$ plane, which is spanned by the vector of the impact
parameter and the beam direction. In the rest source frame, the
three momentum components $P'_x$, $P'_y$, and $P'_z$ of the
considered particle have Gaussian distributions with the same
standard deviation $\sigma$.  The azimuthal anisotropy of particles
produced in nuclear collisions can be described by the two momentum,
$P_x$ and $P_y$ components in the transverse plane. The $P_x$ and
$P_y$ distributions are given by

\bea f_{P'_{x,y}}(P'_{x,y})=\frac{1}{\sqrt{2\pi}\sigma}\exp
\left[-\frac{P'^2_{x,y}}{2\sigma^2}\right]. \eea According to
Gaussian momentum distribution, the width of the distribution
$\sigma=\sqrt{TM}$, where $T$ and $M$ are the source temperature and
particle mass respectively.

 Due
to the interactions with other emission sources, the considered
source will have deformations and movements along the $ox$ and $oy$
axes, we have

 \bse \label{momentum}\bea
P_x=A_xP~'_x+B_x,\\
P_y=A_yP~'_y+B_y,
 \eea\ese
where $A_{x,y}$ and $B_{x,y}$ represent the coefficients of source
deformations along the $ox$(or $oy$) axis and  movements along the
corresponding axis respectively.  The shifted deformable source is
simply parameterized to have the linear relation between $P_{x,y}$
and $P'_{x,y}$, which reflects the mean result of source
interaction. For a Maxwell¡¯s ideal gas, $A_{x,y}=1$ and
$B_{x,y}=0$, the source has no expansion and no contribution to the
collective flow. For $A_{x,y}>1$, the source expands,  leading to an
azimuthal anisotropy of the particle momentum distribution in the
transverse plane. The amplitude of the azimuthal anisotropy can be
measured by the elliptic flow, which is a sensitive probe of the
final state interactions. For $B_{x,y}$, a positive value means
source movements along the direction of the $ox$ or $oy$ axis, and a
negative value indicates source movements along the opposite
direction of the corresponding axis.

Generally speaking, different $A_{x,y}$ or $B_{x,y}$ can be obtained
for events with different centralities or impact parameters. When
two ultrarelativistic nuclei collide at non-zero impact parameter,
their overlap area in the transverse plane has a short axis($ox$),
 which is parallel to the impact parameter, and a long axis which is
  perpendicular to the impact parameter. Elliptic flow in the central
  region of collisions is driven by the
almond shape of the overlapping
region~\cite{Sorge:1996pc,Drescher:2007cd,Snellings:2011sz}. This
almond shape of the initial profile is converted by the pressure
gradient into a momentum asymmetry. More particles are emitted along
the short axis and the magnitude of this effect is characterized by
$A_x$. The momentum anisotropy is largest in the early evolution of
the collision. As the system expanding, the system becomes more
spherical, and the driving force quenches itself gradually. With the
increasing of $P_x$ or $P_y$ of the particles produced in the
collisions, the momentum anisotropy decreases. So, for the concerned
centrality, $A_x$ is not a constant, but a function of the momentum.
We set empirically the function
 \be
A_x(P_x, P_y)=1+k \exp \biggl(-\sqrt{(u_1P_x)^2+(u_2P_y)^2}{\bigg
/}\sigma\biggl)
 \ee
where the expansion factor $k$ indicates the magnitude of the
expansion, $u_1$ and $u_2$ denote the contributions of $P_x$ and
$P_y$ respectively. When $k>0$, we have $A_x>1$, which renders the
expansion of sources along  the axis. When $P_x$(or $P_y$) increases
to sufficiently large values, $A_x$ closes to 1, which renders the
momentum anisotropy vanishes and the elliptic flow $v_2=0$.  The
other parameters taken for calculations are $A_y=1$ and $B_x=B_y=0$,
which is the default value.

The probability density function of $P_{x,y}$ is obtained as \bea
f_{P_{x,y}}(P_{x,y})=\frac{1}{\sqrt{2\pi}\sigma A_{x,y}}
\exp\left[-\frac{(P_{x,y}-B_{x,y})^2}{2\sigma^2A_{x,y}^2}\right] .
 \eea
The corresponding transverse momentum and azimuth read as
 \be
P_T=\sqrt{P_x^2+P_y^2}\ee and \be
\label{phi}\phi=\arctan{\frac{P_y}{P_x}}
 \ee respectively. With Eqs. (\ref{dis}), the joint probability density function of
 $P_T$ and $\phi$ is given by
\bea f_{P_T,\phi}(P_T,\phi)
& = &  P_Tf_{P_T,\phi}(P_T\cos\phi,P_T\sin\phi)\nonumber\\
& = &
\frac{P_T}{2\pi\sigma^2A_xA_y}\exp\left[-\frac{(P_T\cos\phi-B_x)^2}{2\sigma^2A_x^2}
-\frac{(P_T\sin\phi-B_y)^2}{2\sigma^2A_y^2}\right] \eea Considering
the measurement in experiments, the elliptic flow in the model is
defined as \be\label{v2}
v_2(P_T)=\langle\cos(2\phi)\rangle=\frac{\int_0^{2\pi}d\phi\int_{P_T-\frac{1}{2}\Delta
P_T} ^{P_T+\frac{1}{2}\Delta
P_T}\cos(2\phi)f_{P_T,\phi}(P_T,\phi)dP_T}
{\int_0^{2\pi}d\phi\int_{P_T-\frac{1}{2}\Delta
P_T}^{P_T+\frac{1}{2}\Delta
P_T}f_{P_T,\phi}(P_T,\phi)dP_T}\,\,\,,\ee where $\Delta P_T$ is a
given $P_T$ bin.
 In the Monte Carlo calculation, we have

\bea P~'_{x,y}=\sigma\sqrt{-2\ln r_{1,3}}\cos(2\pi r_{2,4})\,\,
 \eea
where $r_1$, $r_2$, $r_3$, and $r_4$ denote
 random numbers in [0,1]. Using Eq. (2), (5), and (9),
 the transverse momentum and azimuthal angle can be represented as
\be P_T=\sqrt{\left[A_x\sigma\sqrt{-2\ln r_1}\cos(2\pi
r_2)+B_x\right]^2+\left[A_y\sigma\sqrt{-2\ln r_3}\cos(2\pi
r_4)+B_y\right]^2}\ee and \be
\label{phi}\phi=\arctan{\frac{A_y\sigma\sqrt{-2\ln r_3}\cos(2\pi
r_4)+B_y}{A_x\sigma\sqrt{-2\ln r_1}\cos(2\pi r_2)+B_x}}\,\,\,\,,
 \ee respectively. The elliptic flow reads
\begin{equation}
v_2=\langle\cos(2\phi)\rangle=\left\langle \frac{P_x^2-P_y^2}{P_x
^2+P_y^2}\right\rangle.
\end{equation}\\

{\section{COMPARISONS WITH EXPERIMENTAL DATA}}

The dependences of elliptic flows $v_2$ on transverse momentum $P_T$
for charged particles produced in Au-Au collisions at
$\sqrt{\mathrm{\it s_{NN}}}=200$ GeV are presented in Fig.1 The
symbols are the experimental data of the STAR Collaboration~
\cite{Adare:2010ux} in eleven  collision centralities. Our results
are shown as the curves. The parameters used for calculations and
the corresponding $\chi^2$ per degree of freedom ($\chi^2/{\rm
dof}$) are given in Table I. In the calculation, the values of
$\sigma$ and $k$ are determined by fitting the model to the data,
the values of  $u_1$ and $u_2$ are taken to be 0.944 and 0.59-0.61
respectively, which is independent on centrality. It is obvious that
$v_2$ increases with $P_T$ in the low $P_T$ region, as predicted by
ideal hydrodynamic calculations. The observed $v_2$ saturates or
decreases in the region of $P_T>2$ GeV/$c$. However, ideal
hydrodynamic model calculations show that $v_2$ increases with $P_T$
in this region~\cite{Huovinen:2006jp}. One can see that the
calculated results are approximately in agreement with the
experimental data on the dependence of $v_2$ on $P_T$ in the whole
observed $P_T$ region for all concerned centralities. The expansion
factor $k$ increases and the momentum distribution width $\sigma$
decreases with the centrality percent. According to
$\sigma=\sqrt{TM}$, we can say that the source temperature $T$
increases with the centrality.

In Fig. 2, we show elliptic flows $v_2$ as a function of the
transverse momentum $P_T$ for charged hadrons produced  in Pb-Pb
collisions at $\sqrt{\mathrm{\it s_{NN}}}=2.76$
TeV~\cite{Aamodt:2010pa}. The symbols are the experimental results
obtained with 4-particle cumulant methods and denoted as $v_2\{4\}$.
The elliptic flow of charged hadrons measured by the ALICE
Collaboration as a function of transverse momentum $v_2(P_T )$ is
nearly identical to that measured by the STAR Collaboration at RHIC
up to $P_T=3$ GeV/$c$, independent of collision energy and
centrality. The calculated results are approximately in agreement
with the experimental data on the dependence of $v_2$ on $P_T$ for
$h^\pm$ in Pb-Pb collisions at $\sqrt{\mathrm{\it s_{NN}}}=2.76$
TeV. In the calculation, we take $u_1=0.988$ and $u_2=0.625-0.640$.
The values of the expansion factor $k$ increase with the centrality
percent, and are systematically larger than those in Au-Au
collisions at $\sqrt{\mathrm{\it s_{NN}}}=200$ GeV. The values
$\sigma$ and $T$ also decreases with centrality percent, as that in
Fig.1.

The square  values of the expansion factor $k^2$ used for Fig.1 and
Fig.2 are displayed in Figs.3, by different symbols as marked. In
Au-Au and Pb-Pb collisions, $v_2(P_T)$ tends to a saturation for the
more peripheral collisions($>40\%$).  We find that, except for the
saturation region, the  values of $k^2$ exhibit a linear dependence
on centrality percent$c$. For Au-Au and Pb-Pb collisions, we have
$k^2=(3.161\pm 0.002)c $ and $k^2=(3.492\pm 0.012)c$, respectively.
By the geometric relation of centrality percent $c$ to the impact
parameter $b$, we have $c\propto b^2$ , which holds to a very high
accuracy for all but most peripheral collisions. Thus,  the
expansion factor $k$ linearly increases with the increasing of the
impact parameter $b$ from most central (0-5\%) to mid-peripheral
(35-40\%) collisions.

 {\section{conclusions }}

In the above discussions, we have investigated the results of
azimuthal anisotropy by using the elliptic flow $v_2$  vs the
transverse momentum $P_T$ for several centralities  in
$\sqrt{\mathrm{\it s_{NN}}}=200$
 GeV Au-Au and $\sqrt{\mathrm{\it s_{NN}}}=2.76$ TeV Pb-Pb
 collisions. The calculated results are obtained  in the framework of
 the multi-source
 idealgas model by comparing the experimental data.
 The calculated results are approximately in
 agreement with the experiment data of the STAR and ALICE
 Collaboration. In our model, the local sources of  final observed particles are
formed in heavy ion collisions. Their interaction are  related to
the hot dense matter in the sources, and also results in  the
azimuthally  anisotropic expansion in the momentum space
\cite{Song:2010er}. The almond shaped interaction volume produced in
a non-cental collision is converted by the pressure gradient into a
momentum asymmetry. The parameter $A_x$ in the model is used to
reflect the expansion of momenta, as shown in Eq.(3). An isotropic
emission corresponds to $k=0$.

 From the above comparisons, we can see that the values of the expansion factor $k$
 increases linearly with the impact parameter $b$ increasing over the range of collision
 centrality (from 0-5\% to 35-40\%) in $\sqrt{\mathrm{\it s_{NN}}}=200$
 GeV Au-Au and $\sqrt{\mathrm{\it s_{NN}}}=2.76$ TeV Pb-Pb
 collisions. It supports the idea that the  increase from central to peripheral collisions
reflects the expected increase due to the change in initial
eccentricity of the  initial fireball from central to peripheral
events at each centrality~\cite{Bhalerao:2005mm,Broniowski:2007ft}.
The difference between the two collisions is that the expansion
factor of the latter one is larger than that of the former one. From
RHIC to LHC energies, it is found that the average QGP specific
shear viscosity $\eta/s$ slightly increases~\cite{Song:2011qa,
Roy:2011xt}. Recent hydrodynamic calculations in the framework of
relativistic dissipative hydrodynamics have been interpreted as the
evidence that the elliptic flow at RHIC energies is insensitive to
the QGP viscosity and only becomes sensitive to it at LHC
energies~\cite{Niemi:2011ix}.

 The azimuthal anisotropy resulting from the final-state
particles is one of the most informative quantities in better
understanding the nature and properties of the matter in high energy
nuclear collisions. The ideal hydrodynamic model calculations
reproduce the mass ordering of $v_2$ in the relatively low $P_T$
region, but overshoot the values of $v_2$ for all centrality
bins~\cite{Abelev:2010tr}. To understand the viscous nature of QGP,
dissipative hydrodynamics has recently been applied to explain the
experimental data of the collective flow parameter $v_2$ by
including the effect of shear and bulk viscosity
~\cite{Denicol:2010tr, Schenke:2011tv,Shen:2011kn}. Analysis of the
elliptic flow in our simple model, where the system expansion can be
quantified in the momentum space, shows that the expansion factor
$k$ is characterized by the impact parameter $b$, which is related
to participating nucleons using a realistic description of the
nuclear geometry in a Glauber calculation~\cite{Miller:2007ri} .

In summary, in the framework of the multi-source idealgas model, we
have investigated the transverse momentum dependence of elliptic
flow for charged particles in Au-Au collisions at the RHIC energy
and in Pb-Pb collisions at the LHC energy. Our calculated results
obtained in the multi-source idealgas model approximately agree with
the experimental data.  The  expansion factor $k$ used in the
calculation exhibits a linear dependence on the impact parameter $b$
from most central (0-5\%) to mid-peripheral (35-40\%) collisions.
Elliptic flow has been proved to be very valuable for understanding
relativistic nuclear collisions. As we know that the hydrodynamic
description has gained important result over the past few years. The
present work shows that the simpler multi-source idealgas model also
can do the similar job. Particularly, in the descriptions of
(pseudo)rapidity and multiplicity distributions for produced
particles, this model is successful. In the description of the
$v_2(P_T)$ at higher  energies, the work  is a successful attempt.

{\bf Acknowledgments.} This work is supported by the National
Natural Science Foundation of China under Grant No. 10975095, the
National Fundamental Fund of Personnel Training Grant No. J0730317
and the Open Research Subject of the CAS Large-Scale Scientific
Facility Grant No. 2060205.

\newpage

\begin{table}[th]
\vspace*{-0.0cm} {\small \vspace*{-0.1cm}
\caption{Values of parameters $k$ and $\sigma$ in our calculations.  }%
\label{table-all}%
\vspace*{-0.15cm}
\begin{center}
{\begin{tabular}{c| c c c c c} \hline\hline \,\,\,\,Figure\,\,\,\,&
\,\,\,\,Centrality
&\,\,\,\, $\sigma$(GeV/c) \,\,\,\,& $k$(GeV/c) & \,\,\,\,$\chi^2$/dof\,\,\,\, \\
\hline
 Fig.1   & 50\%-60\%  & $1.119\pm 0.006$  & $1.076\pm 0.001$ &0.707\\
         & 45\%-50\%  & $1.127\pm 0.004$  & $1.069\pm 0.002$ &0.716\\
         & 40\%-45\%  & $1.141\pm 0.011$  & $1.059\pm 0.005$ &0.643\\
         & 35\%-40\%  & $1.164\pm 0.016$  & $1.046\pm 0.008$ &0.622\\
         & 30\%-35\%  & $1.204\pm 0.024$  & $0.991\pm 0.010$ &0.495\\
         & 25\%-30\%  & $1.262\pm 0.025$  & $0.936\pm 0.012$ &0.413\\
         & 20\%-25\%  & $1.336\pm 0.028$  & $0.875\pm 0.024$ &0.244\\
         & 15\%-20\%  & $1.365\pm 0.020$  & $0.719\pm 0.052$ &0.192\\
         & 10\%-15\%  & $1.402\pm 0.019$  & $0.586\pm 0.045$ &0.165\\
         &  5\%-10\%  & $1.452\pm 0.016$  & $0.426\pm 0.042$ &0.146\\
         &  0\%- 5\%  & $1.551\pm 0.016$  & $0.264\pm 0.036$ &0.104\\
 Fig.2   & 40\%-50\%  & $1.192\pm 0.020$  & $1.116\pm 0.002$ &0.808\\
         & 30\%-40\%  & $1.248\pm 0.022$  & $1.110\pm 0.004$ &0.244\\
         & 20\%-30\%  & $1.344\pm 0.018$  & $0.942\pm 0.056$ &0.124\\
         & 10\%-20\%  & $1.486\pm 0.015$  & $0.725\pm 0.044$ &0.116\\

\hline\hline
\end{tabular}}
\end{center}}

\end{table}

\begin{figure}[h]
\includegraphics[width=0.8\textwidth] {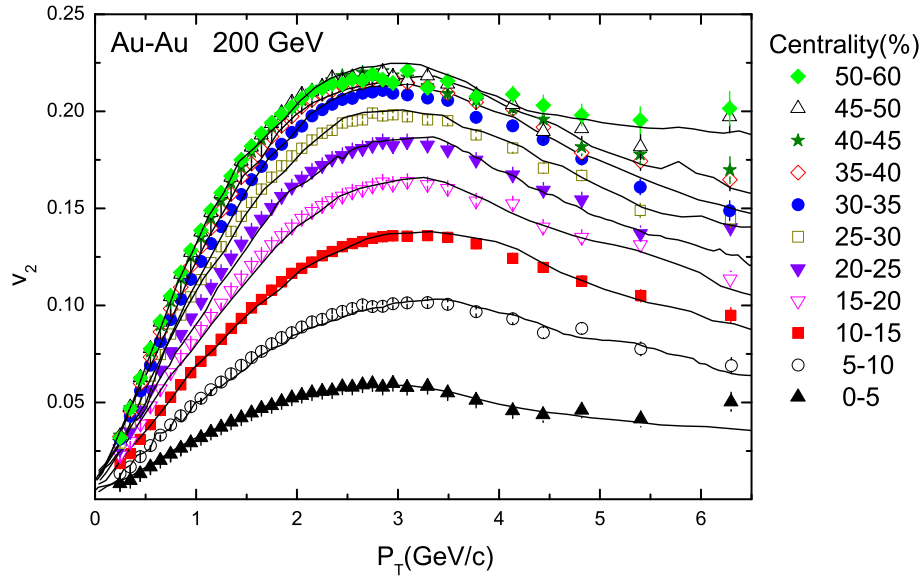}

\caption{$P_T$ dependence of $v_2$ for charge hadrons  in Au-Au
collisions at $\sqrt{\mathrm{\it s_{NN}}}$ = 200 GeV.  Experimental
data taken from the STAR Collaboration~ \cite{Adare:2010ux} are
shown with the scattered symbols. Our results calculated from the
multi-source idealgas model are shown with the curves.}
 \label{S1L}
\end{figure}

\begin{figure}[h]
\includegraphics[width=0.75\textwidth] {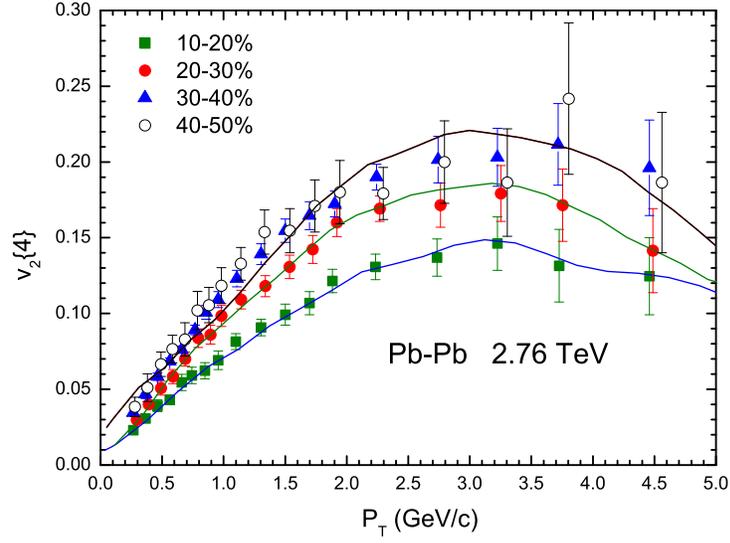}

\caption{$P_T$ dependence of $v_2$ for charge hadrons in Pb-Pb
collisions at $\sqrt{\mathrm{\it s_{NN}}}$ = 2.76 TeV.  Experimental
data taken from the ALICE Collaboration~\cite{Aamodt:2010pa} are
shown with the scattered symbols. Our results calculated from the
multi-source model idealgas are shown with the curves.}
 \label{S2L}
\end{figure}

\newpage
\begin{figure}[b]
\includegraphics[width=0.5\textwidth] {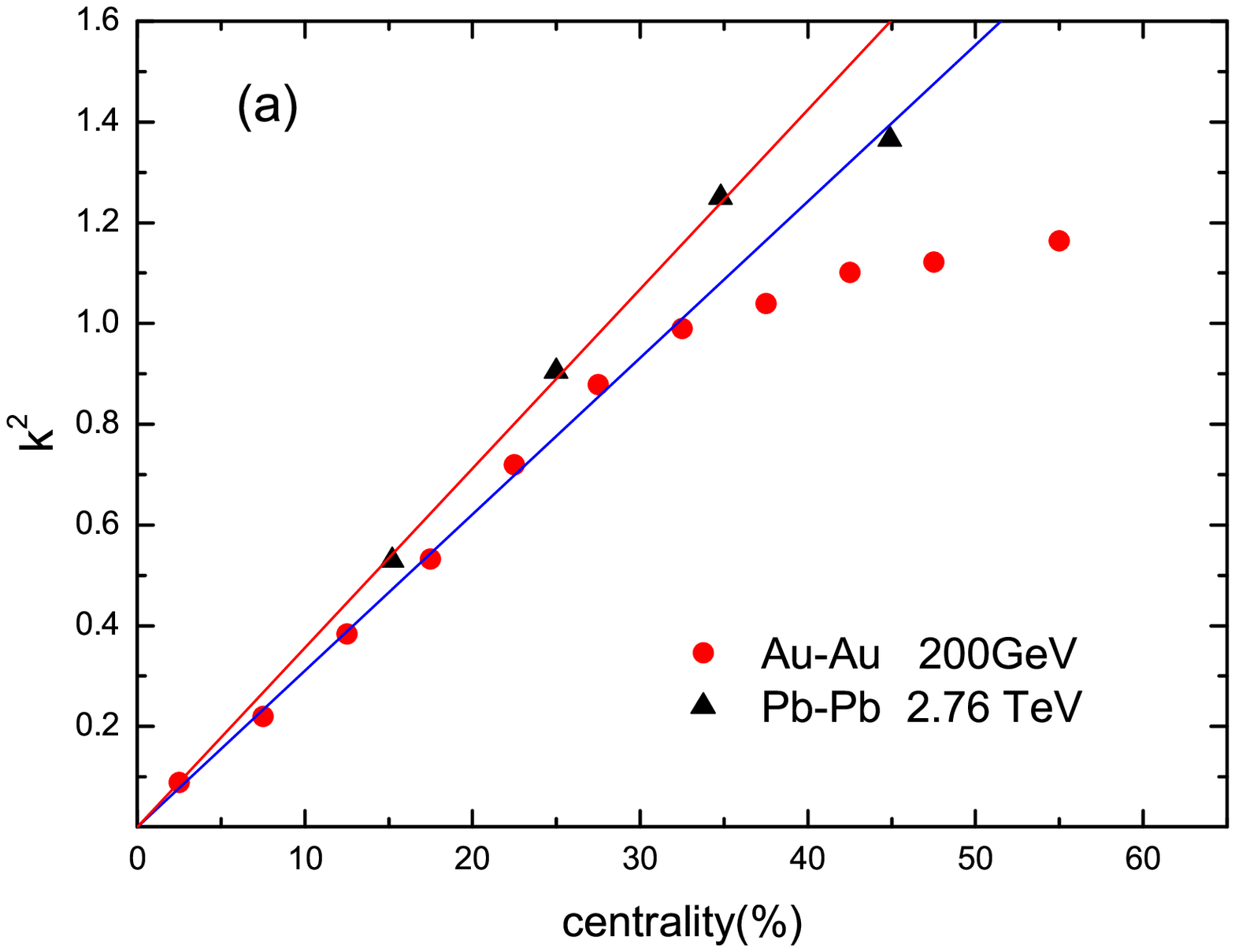}\hspace{-1.0cm}
\includegraphics[width=0.5\textwidth] {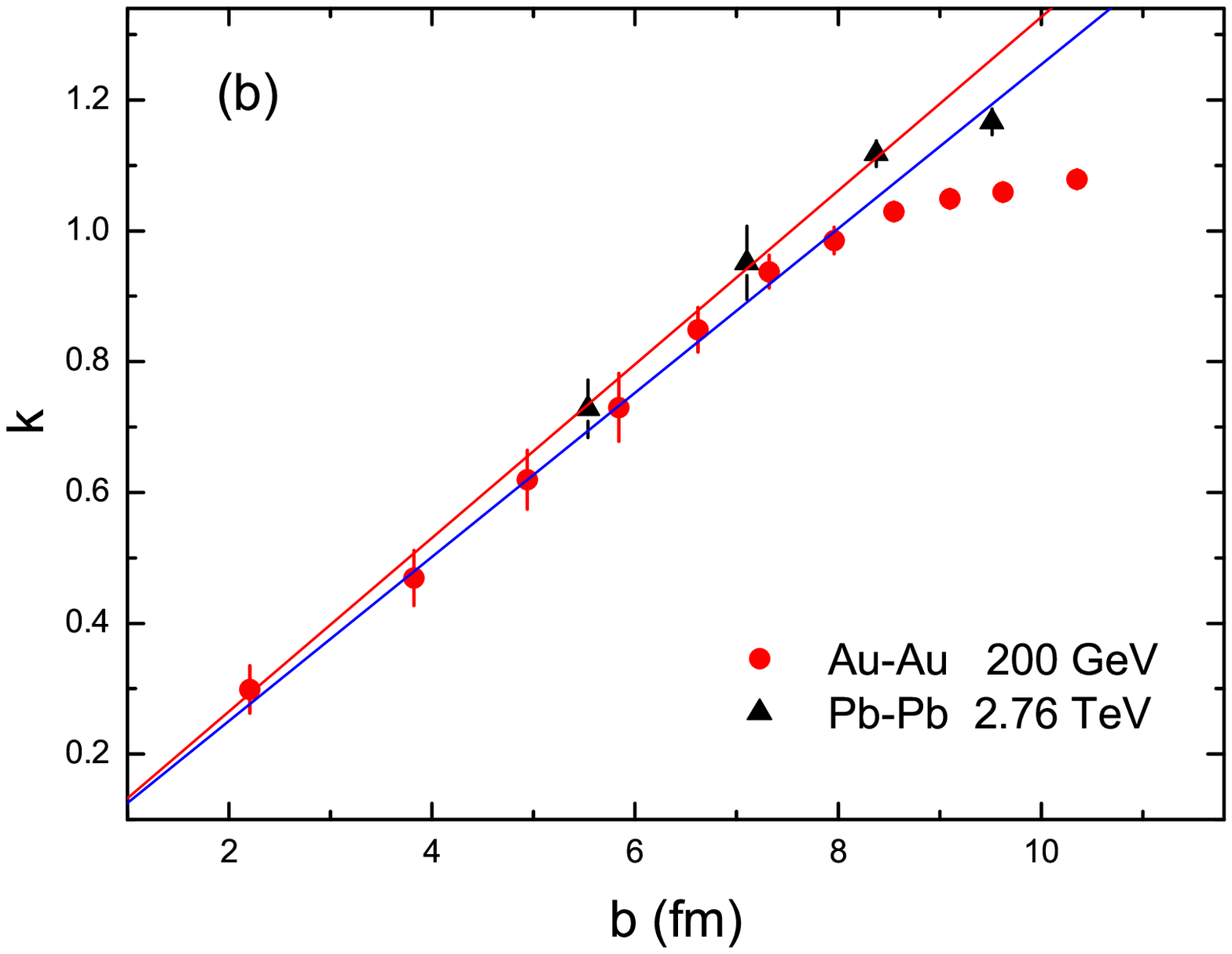}

\caption{Centrality and the impact parameter $b$ dependence of the
square of the expansion factor $k^2$. The symbols represent the
parameter values used in Figs.1-2. The lines are a fitted results.}
 \label{S1L}
\end{figure}\newpage

\end{document}